# Discovery of Convoys in Network Proximity Log

Dmitry Namiot
Lomonosov Moscow State University
Faculty of Computational Math and Cybernetics
Moscow, Russia
e-mail: dnamiot@gmail.com

*Abstract*— This paper describes an algorithm for discovery of convoys in database with proximity log. Traditionally, discovery of convoys covers trajectories databases. This paper presents a model for context-aware browsing application based on the network proximity. Our model uses mobile phone as proximity sensor and proximity data replaces location information. As per our concept, any existing or even especially created wireless network node could be used as presence sensor that can discover access to some dynamic or user-generated content. Content revelation in this model depends on rules based on the proximity. Discovery of convoys in historical user's logs provides a new class of rules for delivering local content to mobile subscribers.

I. INTRODUCTION

The term a trajectory refers to the movement of an object given by a continuous curve in the space. The past trajectory of an object is usually approximated by a collection of time stamped positions. For example, in our research we target mobile phones where positions usually could be obtained from a GPS device.

Convoy is a group of objects that travel together for more than some minimum duration of time. More probably, that the original task for discovery of convoys (groups of objects with coherent trajectory patters) was oriented to the military applications. As per nowadays research papers, a number of applications may be envisioned. The discovery of common routes among citizens may be used for the scheduling of public transport. The discovery of convoys for trucks may be used for throughput planning. The identification of cars that follow the same routes approximately at the same time may be used for creating carpooling, etc.

In our paper we will follow to convoy definitions from [1] and avoid restrictions on the sizes and shapes of the discovered trajectory patterns. Generic trajectory pattern of any shape and any extent will be based on the notion of density connection [2]. It enables the formulation of arbitrary shapes of groups.

Shortly, convoy is a group of moving object where included objects are in density connection the consecutive time points. Objects are density-connected if a sequence of objects exists that connects the two objects and the distance between consecutive objects does not exceed the given value. As it follows from this definition, convoy definition depends on the time, during which the objects in the density-connected group traveled together. As per our target area, we will consider relatively short traveling time and does not consider the distances between pairs of trajectories across all of time.

The next often used in this context terminology is moving cluster (or cluster of moving objects) [3]. The moving cluster exists if a shared set of objects exists across some finite time, but objects may leave and join a cluster during the cluster's life time. So, the semantic is different and moving clusters do not necessarily qualify as convoys (in the pure terms). Both the location and the set of objects of a moving cluster change over time. But sometimes, both definitions are mixed and moving cluster means the same as convoy. Some of authors define dynamic convoys and evolving convoys [4]. Dynamic convoys allows dynamic members under constraints imposed by some parameters (actually, by user-defined parameters). An evolving convoy captures the relationship between different stages of convoys, so that convoys in some stage has more (fewer) members than its previous stage. Another interesting term in this space is flock. Flock is a set of objects that travel within a range while keeping the same motion. Anyway, all patterns covering capturing "collaborative" or "group" behavior between moving objects. The difference between all the above mentioned patterns is the way they define the relationship between the moving objects.

In this paper we will investigate a special case for convoys (moving clusters) discovery. At the first hand, we have not location information in our database. We are working with some context-aware data discovery application, which lets mobile users get hyper-local content right on the mobile phones. This application (namely, SpotEx, first time described in [5]) based on the network proximity. Shortly, it defines logical rules (productions) that depends on the network proximity and lets mobile users discover local content via fired rules. Also, this application can record historical proximity data during the execution. This proximity log becomes the analogue of trajectory database. For example, this application, being executed indoor, records the track (in proximity terms, again) for the mobile user. Discovery of convoys (coherent motions) in such database let us define new class of rules. For example, in proximity marketing applications we can unveil a special kind of offers for those reached our point of sale (be nearby in proximity terms) in the group, etc.

The structure of our proximity log and the way we are getting measurements caused the need for the yet another definition of convoys. It is provided below.

The rest of the paper is organized as follows. Section II contains an analysis of existing approaches for discovery of convoys. It covers, at the first hand, the aspects we need for the future development. In Section III, we describe our SpotEx approach. In Section IV we describe discovery of convoys for our proximity logs.

## II. THE DISCOVERY OF CONVOYS AND NETWORK MEASUREMENTS

This section contains the basic definition for the covered area. Analyzing research papers, we can list the following key issues in the discovery process [6]:

a) cluster related issues. How to define and described cluster for objects?
b) consistency related issues. The detected groups should be consistent enough to last for a given time.
c) group size related issues. Many applied tasks may have requirements on the cluster's size.

Let us give the basic definitions for this area.
Neighborhood: given a distance threshold $e$ and a set of points $S$, distance operator D the e-neighborhood of a point $p$ is given as $NH_e(p) = \{q \in S \mid D(p, q) \leq e\}$.
Density-reach: given a distance threshold $e$ and an integer $m$, a point $p$ is directly density-reachable from a point $q$ if $p \in 2\ NH_e(q)$ and $\mid NH_e(q)\mid \geq m$. A point $p$ is density-reachable from a point $q$ with respect to $e$ and $m$ if there exists a chain of points $p_1, p_2, \ldots, p_n$ in set $S$ such that $p_1 = q$, $p_n = p$, and $p_{i+1}$ is directly density-reachable from $p_i$.
Density connection: given a set of points $S$, a point $p \in S$ is density-connected to a point $q \in S$ with respect to $e$ and $m$ if there exists a point $x \in S$ such that both $p$ and $q$ are density-reachable from $x$.

The definition of density-connected elements is the basic formation for the definition of convoys. Figure 1 illustrates this.

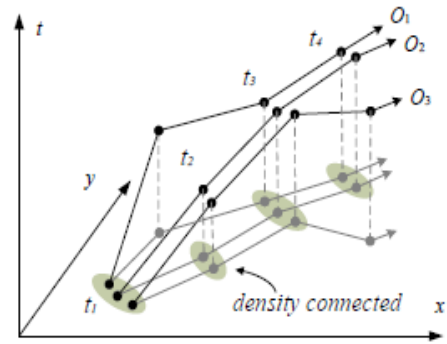

Figure 1. An example of convoy [1]

Given the density-connected objects for consecutive time points, the convoy could be defined as follows: given a set of trajectories of $N$ objects, a distance threshold $e$, an integer $m$, and an integer lifetime $t$, the convoy query returns all possible groups of objects, so that each group consists of a (maximal) set of density-connected objects with respect to $e$ and $m$ during at least $t$ consecutive time points [1].

Another definition, which could be more interesting to our future development, is traveling company [6]. A group of objects form a traveling company, if members of the group are density-connected for themselves during some given time and the size of the group is not less than the given threshold.

In order to discover the traveling groups, most of the algorithms use the concepts of density-based clustering [7]. The simplest and, probably, most often used technique for discovering of convoy is to perform density-connected clustering on the objects at each time and then extract their common objects. Note, that our trajectories may have some missing time points due to non-regular sampling for locations. It prevents us from the checking the density-connection for all objects involved over those missing times. Some authors suggest linear interpolation for creating virtual points for missed times (e.g., Coherent Moving Cluster algorithm in [1]).

An idea to use network measurements for moving detection has been presented in many papers. For example, Locadio [8] uses Wi-Fi signal strengths from existing access points measured on the client to infer both pieces of context. For motion, authors measure the variance of the signal strength of the strongest access point as input to a simple two-state hidden Markov model (HMM) for smoothing transitions between the inferred states of "still" and "moving." This was based on the observation that when a Wi-Fi receiver (mobile phone) is moving, the signal strengths it receives are noisier than when it is not moving. For location, authors exploit the fact that Wi-Fi signal strengths vary with location.

Software based systems that provide location estimation based on the received signal strength indication (RSSI) of wireless access points are becoming popular nowadays. The main benefit of RSSI measurement based systems is that they do not require any additional sensors or actuators and can use existing infrastructure and already available communication parameters.

In general, RSSI based positioning includes two phases:
- the training phase where the wireless map of the environment is determined using field measurements
- the positioning phase where position estimation is performed based on the wireless map. Note that the training phase is an offline process and as such only needs to be redone if there have been major changes to the wireless propagation environment (e.g., relocation of access points) [9]

Note, the calibration phase could be very costly actually. Also, it does not support dynamic Wi-Fi nodes. What can we do if our Wi-Fi hot spot will be opened right on the mobile phones?

Technically, RSSI based measurement approaches can be divided into deterministic and probabilistic techniques.

In deterministic techniques our location area is subdivided into smaller cells and training phase readings are taken in these cells from several known access points. In the positioning phase the most likely cell (the cell that best fits the current measurement) is selected.

In probabilistic positioning techniques a probability distribution of the user's location is defined over the area of the movement. The goal of the positioning is to reach a single mode for this distribution, which is the most likely location of the tracked user. Probabilistic approaches to mobile node positioning from RSSI measurements rely on the precise estimation of a posterior probability distribution, $p(s_t \mid d_1, \ldots, d_t)$, of the likelihood of the node's state (location), $s_t$, given a history of the received measurements, $d_1, \ldots, d_t$ [9].

### III. SPOT EXPERT AS A NETWORK PROXIMITY SERVICE

Originally, the main idea of Spot Expert (SpotEx) comes as an extension for Wi-Fi based indoor positioning service (IPS). Spotex uses only a part of Wi-Fi based IPS. It stops process on the phase of detection Wi-Fi networks. Due to local nature of radio interfaces in Wi-Fi, this detection already provides some information about the location. More precisely, we can get information about proximity. As the second step, we add an external database with some rules (productions or if-then operators), related to the Wi-Fi access points. The typical examples for conditions in our rules are: Access point with SSID SomeCafé is visible for mobile device; time is within the given interval, signal strength is within the given interval, etc. And based on such conclusions, we will provide context-aware data retrieval and present some user-defined messages to mobile terminals. In other words, in SpotEx content's visibility depends on the network context (fingerprint for Wi-Fi network environment).

For the first time, SpotEx service [10], developed by Dmitry Namiot was described in article published in NGMAST-2011 proceedings [5]. You can see the latest state of SpotEx development in papers [11] and [12], for example.

SpotEx model does not require calibration phase as the most Wi-Fi based IPS do and based on the ideas of network proximity. Proximity based rules replace location information, where Wi-Fi hot spots work as presence sensors. SpotEx approach does not require from mobile users to be connected to the detected networks. SpotEx uses only broadcasted SSID for networks and any other public information.

Technically, SpotEx contains the following components:

- Server side infrastructure. It includes a database (store) with productions (rules), rules engine and rules editor. Rules engine is responsible for runtime data retrieval. Rule editor is a web application that lets work with rules database.

- Mobile application. This part is responsible for getting context info, matching it against productions (rules) and visualizing the output

SpotEx could be deployed on any existing Wi-Fi network (or networks especially created for this service) without any changes in the infrastructure. Rule editor lets easily define some rules described context visibility to that network. Context here is just some text (HTML code) that should be opened (delivered) to the end-user's mobile terminal as soon as the appropriate rule is fired. For example, as soon as one of the above-mentioned networks is getting detected via our mobile application.

Existing use cases target proximity marketing, at the first hand. The whole process looks like an "automatic check-in" (by analogue with Foursquare, etc.) One shop can deliver proximity marketing materials right to mobile terminals as soon as the user is near some selected access point. Rather than directly (manually or via some API) check-in at the particular place (e.g., similar to Foursquare, Facebook Places, etc.) and get back deals info, with SpotEx mobile subscriber can collect deals info automatically. The prospect areas, by our opinion, are information systems for campuses and hyper local news delivery in Smart City projects. Rules could be easily linked to the public available networks.

One interesting use case could be based on the fact that most of the smart phones let users open Wi-Fi hot spots right on the phone. We can associate our rules with such mobile hot spot (hot spots). Another example of mobile hot spot is connected car. In this case our content becomes linked to the phones. It is a typical dynamic LBS. The available services are moving when phone is moving and hot spot is switched on/off. Services automatically follow to the phone.

Smart phone is all what we need for creating a new information channel. It is infrastructure-less approach.

This approach does not discuss security and connectivity issues. We do not need to connect mobile subscribers to our hot spot. SpotEx is all about using hot spot attributes as triggers in data discovery.

Each rule is a logical production (if-then operator). The conditional part includes the following data measured by the mobile application:

Wi-Fi network (SSID, mac-address)
RSSI (signal strength - optionally)
Time of the day (optionally)
ID for the client (mac-address)

In other words it is a set of operators like:

IF IS_VISIBLE('mycafe') AND FIRST_VISIT() THEN {present the coupon info }.

Figure 2 presents use case for proximity marketing in retail area:

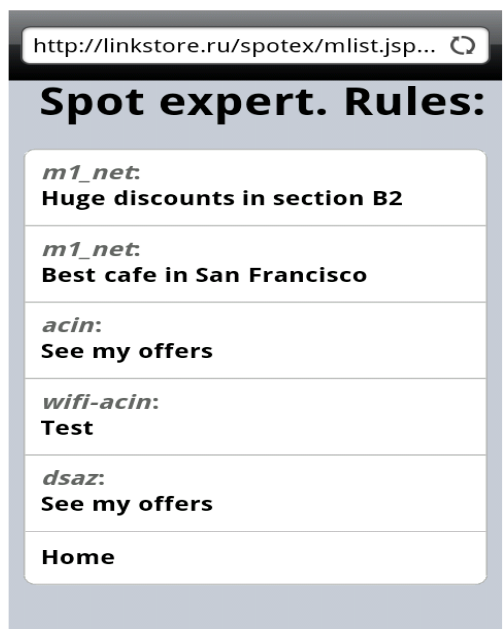

Figure 2.        SpotEx console snapshot

Because our rules present the standard production rule based system, we can use an old and well know Rete algorithm [13] for the processing.

Each rule looks like a production (if-then operator). The conditional part depends on the above mentioned measurements and logical functions (predicates). The predicates (in the current version) are:

```
IS_VISIBLE ( )
NOT_VISIBLE ( )
CLOSE_THAN( )
FIRST_VISIT( )
FOLLOW_UP_VISIT( )

TIME()
TIME_WITHIN()
```

Function IS_VISIBLE() or NOT_VISIBLE() accept as a parameter network ID (e.g., SSID or mac-address for access point) and returns a Boolean value depends on the current network's visibility.

Function CLOSE_THAN() accepts two parameters identified wireless networks (Wi-Fi access points) and returns Boolean value true if mobile terminal is close to the Wi-Fi access point described in the first parameter.

Two functions FIRST_VISIT() and FOLLOW_UP_VISIT( ) based on the simply fact that in Wi-Fi based system we have MAC-address for mobile terminal. The whole system does not require authorization. With SpotEx users can discover data anonymously. But in the same time we have some analogue of UUID, allowing us distinguish the users. It is MAC-address. We keep historical logs for vectors (MAC-address, wireless environment info) and use it for detecting new or retuned "visitors". For example, if for the same MAC-address we have at least two historical records where at least one Wi-Fi access point mentioned twice or more it is follow-up visitor.

Here is the starting point for our discovery of trajectories. Via the recorded track of networks snapshot environment we can try to discover how the current point (moment, when we are checking rules for the particular user) was reached. It let us define rules that depend on that track.

Let us describe another example. In the modern LBS applications that are mostly circling near the idea of "check-in", we lack the history of the movement almost completely. It is especially true on the micro-level (indoor). Suppose I have a new check-in in Foursquare. How do I come to this place? In the ordinary web browsing, any hyperlink click can have a referrer field. There are no references in LBS. In this paper we present our initial attempt to fill this gap. On practice, it means that we are going to add to our predicates a new logical function:

IN_GROUP_OF (n, t)

Here $n$ presents some positive integer value and $t$ describes a time (e.g., seconds). This function returns Boolean value *true* if mobile user traveled in the group of at least $n$ people during at least $t$ seconds. It is, by the SpotEx vision, of course, and those $n$ people should be presented via own records in the proximity log. We think, that such a function (actually – qualification for context) could be useful in proximity marketing tasks, for data discovery in Smart City projects, etc.

For example, SpotEx supports an external database for customized check-ins. Any such checked "location" is actually some Wi-Fi fingerprint (it is illustrated on Figure 3):

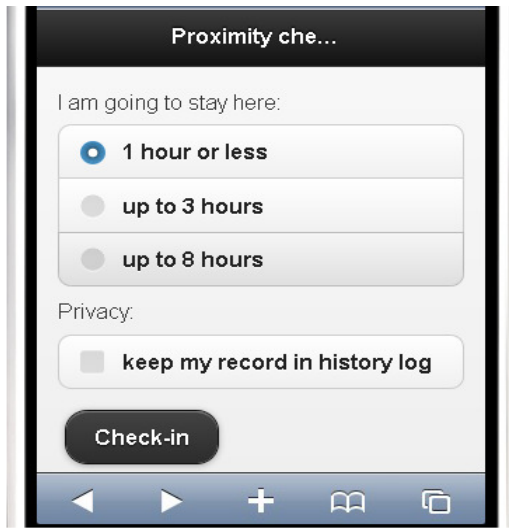

Figure 3. In proximity check-in

This external database just keeps a temporal mapping between IDs in social networks (e.g. Facebook ID) and Wi-Fi fingerprints. And check-ins could be customized, as it is described in [14], for example. In this case, the above-mentioned function IN_GROUP_OF() is a way to present, for example, some special offers to visitors. Group discount in retail is the simplest and obvious use case.

Our discovery process will use recorded wireless network environment snapshots (Wi-Fi fingerprints, actually). SpotEx application collects them from the moment user started the application. Of course, we can investigate historical logs too, but it is separate task. It is something that described in Reality Mining projects [15], for example. In the classical paper, authors perform cluster analysis for the previously collected data. A Hidden Markov Model conditioned on both the hour of day as well as weekday or weekend provided data separation for behavior patterns like "hone", "office", etc.

Of course, SpotEx is not the only approach uses phone as a sensor concept. We've tested the ability to implement our approach with the project Funf [16], for example. Funf Probes are the basic collection data objects used by the Funf framework. Each probe is responsible for collecting a specific type of information. These include data collected by on-phone sensors, like accelerometer or GPS location scans, etc. Actually, in Funf many other types of data (context info) can be collected through the phone. In other words, Funf is a rich data logger. We need only small part of it – collect information about wireless environment. And that log could be a source for data discovery too.

IV. TRAJECTORIES IN THE PROXIMITY LOG

So, our context-aware browser collects wireless networks info during the execution. More specifically, our application collects snapshots that describe current Wi-Fi environment. This environment (it is an analogue of fingerprints used in Wi-Fi based indoor positioning) is a time stamped list of records. Each environment's record is a vector of triples. Each triple describes one Wi-Fi network:

Network ID (SSID)
mac-address
signal strength (RSSI)

and the whole environment could be described as a vector of triples:

$E = \{T_1, T_2, \ldots, T_n\}$

Our fingerprint is just a time stamped environment: [ti, Ei].

So, finally, we have a sequence of time stamped environment records. Technically our recording software (based on SpotEx or Funf) obtains data with regular time intervals. But technically again, not all our data could be available for the processing at the time of the calculation. For saving battery at the first hand, recorder can cache data and update central store in the batches (e.g., for every second, third, etc. cycle). This conclusion raises the important question about missing data. It is a common problem for discovery of trajectories. Of course, a robust system should be tolerant to such cases. There are several approaches for dealing with this problem. One possible solution is based on the introduced inactive period for candidates. This inactivity period is a threshold for the maximal allowed time interval between two position reports of the object. For any object is missing in a snapshot, as long as the inactive period is less than the selected inactivity threshold, the system still assumes that the object is traveling together with the companion in previous snapshot. In our current implementation the missing values simply ignored and appropriate object is deleted from snapshot candidate. Different strategies as well as their robustness are subject for future research.

The second important moment belongs to the general principles of our measurement. Suppose we have Wi-Fi access point with omni-directional antenna. As it is illustrated on Figure 4, having only proximity info we cannot distinguish two groups that actually reached our access point from the opposite directions.

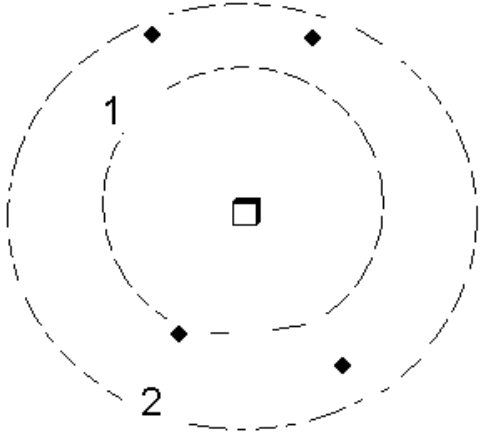

Figure 4.   Omni-directional Wi-Fi antenna

In this case, from the proximity point of view, groups 1 and 2 could be described (detected) as together moved objects. It means that in this paper we will use own definition for traveling groups. For our research it is a group of objects (mobile phones in this particular case) with the similar proximity track within the given time interval. It is consistent movement where the key metric is the relative proximity of an access point. In our research two proximity tracks (sequences of proximity records) are similar on some time interval if for the each sequential measurement in the first track we can get a sequential measurement from the second track for approximately the same timestamp where two networks snapshots have at least one pair of comparable Wi-Fi measurements. Lets us see some example. Suppose we have two tracks:

$T_1$ as $\{[t_{11}, E_{11}], [t_{12}, E_{12}], [t_{13}, E_{13}], \ldots\}$
and
$T_2$ as $\{[t_{21}, E_{21}], [t_{22}, E_{22}], [t_{23}, E_{23}], \ldots\}$

Here $t_{ij}$ describes a time stamp and $E_{ij}$ describes Wi-Fi environment. The similarity means that we can map measurements from the first track to the second one. And our mapping should keep the time sequence. So, for example, if we map a pair $[t_{11}, E_{11}]$ to $\{[t_{21}, E_{21}]$, then the next pair $[t_{12}, E_{12}]$ could be mapped to the time $t \geq t_{21}$.

Because each application (each mobile phone) executes and collects data independently, we can not warranty that for the given timestamp $t_{1i}$ we will find exactly the same value $t_{2j}$ in the second track. We will try to find approximately the same timestamp $t_{1i} \pm \Delta$ where $\Delta$ is some constantly selected threshold. Of course, it could be selected accordingly to the regular interval used for collecting measurements. In other words, comparing to the traditional trajectories discovery algorithms, this application does not restore (does not approximate or predict) missed values. If we cannot find some measurement within the given interval than we simply conclude that two tracks are not similar.

Citing absolutely the same reason (each phone collects data without the correlation with other possible participants) we can not warranty too, that two mobile phones record equal values for Wi-Fi signal strength on any selected place. It depends on battery level, external conditions, etc. It means that we will compare network measurements using the second given threshold (for signal strength). IDs for access points should be equal of course, where RSSI may vary within the given interval. That is why we mentioned above the comparable (approximately equal) Wi-Fi measurements.

Two networks measurements are comparable in this paper (it is the simplest metric) if they have at least one common access point with difference in signal strengths less than the given threshold.

Now we are ready to present our algorithm. We are calculating the Boolean value for function

IN_GROUP_OF( )

This function will be calculated in some of our predicates checked for the given mobile user. It means, that as a starting point we have data for the current wireless environment (Wi-Fi environment snapshot for this mobile user / mobile phone). Initial parameters are:

$\Delta$ - time threshold, $\Omega$ - RSSI threshold, $E$ - an original network environment, $T_0$ – an original current time, $T_{max}$ – argument for function

1. Initialize new candidate set $R_1$
2. Collect measurements within the time $T_0$-$\Delta$ $\rightarrow$ $R_1$;
3. **If** $R_1$ is empty **then** output *false;*
4. Remove from $R_1$ all measurements that are not comparable with $E$;
5. **If** $R_1$ is empty **then** output *false;*
6. **Set** $t = T_0$;
7. **While** $t > T_0$-$T_{max}$
    8. Find the previous measurement for the current user. Update current settings $\rightarrow \{t, E\}$;
    9. For the each measurement in $R_1$ find proximity data within $t \pm \Delta$ (update measurements with new data);
    10. Remove from $R_1$ elements without new data (not updated elements) ;
    11. Remove from $R_1$ elements that are not comparable with $E$;
    12. **If** $R_1$ is empty **then** break;
13. **End while**

The finally, $R_1$ presents the group we are looking for. Depends on the size of this array, we can calculate our function IN_GROUP_OF( ).

## V. CONCLUSION

This article describes a new application (a new use case) for discovery of convoy task. This is an attempt to apply the known models for new use cases generated by the context-aware computing. Network proximity log is used here as a replacement for the classical trajectory database. This fact, as well as the technical aspects of how measurements are collecting, requires changes in the standard definitions and the corresponding modifications for the algorithms. The results of this research will be used for extending the functionality of a new service for context-aware data discovery. The future developments will include the analysis of the stability for the proposed algorithm to missing values and obtaining quantitative metrics for the speed and accuracy of recognition.